\documentclass[twocolumn]{aastex631}

\newcommand{\psr}{PSR~J1537$+$1155}
\newcommand{\psrb}{PSR~B1534$+$12}
\newcommand{\maspy}{$\rm mas~yr^{-1}$}
\newcommand{\kmps}{$\rm km~s^{-1}$}

\newcommand{\psrpi}{\ensuremath{\mathrm{PSR}\pi}}
\newcommand{\mspsrpi}{\ensuremath{\mathrm{MSPSR}\pi}}

\newcommand{\rcs}{$\chi_{\nu}^{2}$}
\newcommand{\fsps}{$\rm fs~s^{-1}$}

\newcommand{\multilinecomment}[1]{}

\shorttitle{VLBA astrometry of \psr}
\shortauthors{Ding et al.}
\graphicspath{{./}{figures/}}

\begin{document}

\title{The orbital-decay test of general relativity to the 2\% level with 6-year VLBA astrometry of the double neutron star \psr}

\author[0000-0002-9174-638X]{Hao Ding}
\affiliation{Centre for Astrophysics and Supercomputing, Swinburne University of Technology, John St., Hawthorn, VIC 3122, Australia}
\affiliation{ARC Centre of Excellence for Gravitational Wave Discovery (OzGrav)}

\author[0000-0001-9434-3837]{Adam T. Deller}
\affiliation{Centre for Astrophysics and Supercomputing, Swinburne University of Technology, John St., Hawthorn, VIC 3122, Australia}
\affiliation{ARC Centre of Excellence for Gravitational Wave Discovery (OzGrav)}


\author[0000-0001-8384-5049]{Emmanuel Fonseca}
\affiliation{Department of Physics and Astronomy, West Virginia University, PO Box 6315, Morgantown, WV 26506, USA}
\affiliation{Center for Gravitational Waves and Cosmology, West Virginia University, Chestnut Ridge Research Building, Morgantown, WV 26505, USA}

\author[0000-0001-9784-8670]{Ingrid H. Stairs}
\affiliation{Dept. of Physics and Astronomy, University of British Columbia, 6224 Agricultural Road, Vancouver, BC, V6J 2B1, Canada}

\author{Benjamin Stappers}
\affil{Jodrell Bank Centre for Astrophysics, University of Manchester, Alan Turing Building, Manchester M13 9PL, UK}

\author{Andrew Lyne}
\affil{Jodrell Bank Centre for Astrophysics, University of Manchester, Alan Turing Building, Manchester M13 9PL, UK}



\begin{abstract}

\psr, also known as \psrb, is the second discovered double neutron star (DNS) binary. More than 20 years of timing observations of \psr\ have offered some of the most precise tests of general relativity (GR) in the strong-field regime. 
As one of these tests, the gravitational-wave emission predicted by GR has been probed with the significant orbital decay ($\dot{P}_\mathrm{b}$) of \psr.
However, compared to most GR tests provided with the post-Keplerian parameters, the orbital-decay test was lagging behind in terms of both precision and consistency with GR, limited by the uncertain distance of \psr. 
With an astrometric campaign spanning 6 years using the Very Long Baseline Array, we measured an annual geometric parallax of $1.063\pm0.075$\,mas for \psr, corresponding to a distance of $0.94^{+0.07}_{-0.06}$\,kpc. This is the most tightly-constrained model-independent distance achieved for a DNS to date. After obtaining $\dot{P}_\mathrm{b}^\mathrm{Gal}$ (i.e., the orbital decay caused by Galactic gravitational potential) with a combination of 4 Galactic mass distribution models, we updated the ratio of the observed intrinsic orbital decay to the GR prediction to $0.977\pm0.020$, three times more precise than the previous orbital-decay test ($0.91\pm0.06$) made with \psr.

\end{abstract}



\section{Introduction}
\label{sec:intro}

\subsection{Pulsars in double neutron star systems}
\label{subsec:DNS_pulsars}
Double neutron stars (DNSs) are prized testbeds on which to evaluate theories of gravity and to probe the composition of neutron stars (NSs).
The DNS merger event GW170817 has been recorded both by gravitational-wave (GW) observatories and electromagnetically \citep[e.g.][]{Abbott17,Abbott17a,Goldstein17,Mooley18}, providing constraints on the interior composition of NSs \citep[e.g.][]{Annala18}. 
The same merger event also strengthens the belief that short Gamma Ray Bursts (SGRBs) are generated by DNS mergers \citep[e.g.][]{Coward12}, though most SGRBs are well beyond the horizon of the current ground-based GW detectors.
In addition, DNS mergers are considered the prime sources of \textit{r}-process elements \citep{Eichler89,Korobkin12,Drout17}. 
To test the connection between DNS mergers and the observed abundance of \textit{r}-process elements in the local universe, an estimate of the DNS merger rate is required, which can be constrained with observations of the Galactic DNS population \citep[e.g.][]{Kim15,Pol19}.

During their steady inspiral stage, DNS systems can be studied by measuring and modeling the pulse time-of-arrivals (ToAs) from a pulsar residing in a DNS system (hereafter referred to as a ``DNS pulsar"). 
So far, 16 known DNS pulsars and 3 suspected ones have been discovered from pulsar surveys (see Table~1 of \citealp{Andrews19}), including two found in globular clusters. 
Though in shallower gravitational potentials compared to DNS mergers, DNS pulsars provide some of the most precise tests on gravitational theories in the strong-field regime with long-term timing observations \citep[e.g.][]{Stairs03,Kramer06,Deller09}.
Gravitational theories are tested with DNS pulsars by comparing observed post-Keplerian (PK) parameters, which quantify effects beyond a simple Keplerian model of motion, to the predictions of a specific gravitational theory, e.g., the general theory of relativity (GR). However, the theory-dependent prediction of each PK parameters relies on the masses of the two DNS constituents. Therefore, one needs at least three PK measurements to test a gravitational theory, as two of them have to be used to determine the two DNS constituent masses (based on the theory to be tested).
The PK parameters include (but are not limited to) $\dot{P}_\mathrm{b}$, $\dot{\omega}$, $\gamma$, $r$ and $s$, which stand for, respectively, the orbital decay, the advance of periastron longitude, the Doppler coefficient, the ``range'' of the Shapiro delay effect and the ``shape'' of the Shapiro delay effect. 
To date, the best test of GR was provided with the double pulsar system PSR~J0737$-$3039A/B \citep{Kramer06}, thanks to the extra independent mass ratio constraint (unavailable for other DNSs).

\subsection{PSR~J1537$+$1155}
\label{subsec:J1537}
\psr\ (also known as \psrb, hereafter referred to as J1537) is the second DNS system discovered \citep{Wolszczan91} in a 10.1-hr orbit.
J1537 shows an exceptionally high proper motion among DNS pulsars (see Table~3 of \citealp{Tauris17}), which has been explained with an unusually large kick of 175--300\,\kmps\ received from the second supernova (in the evolution history of J1537) \citep{Tauris17}. 
Based on the timing observations of J1537, the combined $\dot{\omega}$ -- $\gamma$ -- $s$ test returned consistency with GR at the 0.17\% level \citep{Fonseca14}. However, its observed intrinsic $\dot{P}_\mathrm{b}$ deviated from GR prediction, a result which was thought to be due partly or mostly to the poorly constrained distance to the pulsar \citep{Stairs02,Fonseca14}. 
Furthermore, due to the exceptionally high proper motion, the large uncertainty in the distance to J1537 has become the primary limiting factor of the $\dot{P}_\mathrm{b}$ test (\citealp{Stairs98,Stairs02,Fonseca14}, also explained in Section~\ref{sec:Pbdot_test}).

The hitherto most precise distance to J1537 is $1.051\pm0.005$\,kpc \citep{Fonseca14}, obtained by solving for the distance that matches the orbital period derivative observed with pulsar timing, assuming the correctness of GR \citep{Bell96}. However, such ``timing kinematic distances'' (which, in case of confusion, are conceptually different from the distances derived with radial velocities and a Galactic rotation model, e.g. \citealp{Kuchar94,Wenger18}) cannot be used to test theories of gravity, as GR has been assumed to be correct. 
To carry out the $\dot{P}_\mathrm{b}$ test of GR with J1537, one has to have an independent measurement of its distance \citep{Stairs02} in order to correct the distance-dependent terms from the observed orbital decay. 
Prior to this work, the best independent distance for J1537 has been based on its dispersion measurement (DM) along with a model of the distribution of Galactic free-electron density $n_e$, i.e., $0.7\pm0.2$\,kpc with the TC93 model \citep{Taylor93}. However, there are significant downsides with employing DM-based distances for this purpose.
While generally reliable for the population as a whole, DM-based distances can be inaccurate for individual sources \citep[e.g.][]{Deller09a}. This inaccuracy is more likely for sources at high Galactic latitudes $b$, such as J1537 at $b=48\degr$, due to sparser pulsars (that allow DM measurements) in those directions. Moreover, the two more recent $n_e$ models (NE2001 and YMW16, \citealp{Cordes02,Yao17}) have been built using timing-derived distance of J1537, meaning the DM distance of J1537 is no longer independent for these two $n_e$ models.

Compared to the aforementioned ways to measure the distance to J1537, geometric measurements of the distance to J1537 (based on the change in angle or relative distance to the source as the Earth orbits the Sun) offer the ability to measure the source distance to higher precision and free of model dependency.
Such geometric measurements can be realized with global fitting from pulsar timing or VLBI (very long baseline interferometry) observations in the radio band. Based on pulsar timing, the (timing) parallax of J1537 was measured to be $0.86\pm0.18$\,mas \citep{Fonseca14}.
However, as is pointed out in \citet{Fonseca14}, precise determination of timing parallax is often hampered by the covariance between parallax and DM; the stochastic variations in the latter introduced by the changing sightline between the pulsar and Earth can corrupt the timing parallax. Therefore, VLBI astrometry remains the best way to obtain the most precise model-independent geometric distance to J1537. 

In this letter, we present the astrometric results of J1537 obtained with VLBI observations spanning 6 years. Based on the new distance, we strengthen the $\dot{P}_\mathrm{b}$ test of GR with J1537.
Throughout this letter, the uncertainties are provided at 68\% confidence level unless otherwise stated.

\section{Observations and data reduction}
\label{sec:observations}
As part of the \mspsrpi\ program \citep[e.g.][]{Ding20}, J1537 was first observed with the \textit{Very Long Baseline Array} (VLBA) at around 1.5\,GHz from July 2015 to July 2017, which include 2 2-hr pilot observations under the project code BD179 and 9 1-hr observations under the project code BD192. The astrometric campaign was extended with 6 2-hr VLBA observations between August 2020 and July 2021 under the project code BD229. 
The observation and correlation strategy is identical to that of the \psrpi\ program \citep{Deller19}. ICRF~J150424.9$+$102939 and ICRF~J154049.4$+$144745 were observed as the band-pass calibrator and the primary phase calibrator, respectively.
FIRST~J153746.2$+$114215, 16\farcm3 away from J1537, has been identified and adopted as the secondary phase calibrator. 
At correlation of each observation, pulsar gating, based on pulse ephemerides of J1537 monitored with our timing observations, was applied to increase the S/N of detection.

All correlated data were reduced with the {\tt psrvlbireduce} (\url{https://github.com/dingswin/psrvlbireduce}) pipeline written in {\tt ParselTongue} \citep{Kettenis06}, which bridges python users to the two data-reduction packages {\tt AIPS} \citep{Greisen03} and {\tt DIFMAP} \citep{Shepherd94}. The final image-plane models of the two phase calibrators used for the data reduction can be found at \url{https://github.com/dingswin/calibrator_models_for_astrometry}.

The turbulent ionised interstellar medium between the Earth and J1537 leads to diffractive and refractive interstellar scintillation, which can increase or decrease the pulsar flux density \citep{Stairs02}. In four of the 17 VLBA epochs, scintillation reduced the brightness of J1537 below the detection threshold.
For each of the remaining 13 epochs of detection, the pulsar position and its statistical (or random) uncertainty was obtained by fitting an elliptical gaussian to the deconvolved pulsar image. The acquired pulsar positions are provided in Table~\ref{tab:pulsar_positions}.

\begin{table*}
    \raggedright
    \caption{\psr\ positions in reference to  FIRST~J153746.2$+$114215}
     
    	\begin{tabular}{cccccc} 
		\hline
	project & obs. date & $\alpha_\mathrm{J2000}$ (RA.)  & $\delta_\mathrm{J2000}$ (Decl.)  & $(S/N)_\mathrm{J1537}$ $^\mathrm{b}$ & $(S/N)_\mathrm{SC}$ $^\mathrm{b}$ \\
	code & (yr) & & & &  \\
		\hline
	bd179f0	& 2015.5153 & $15^{\rm h}37^{\rm m}09\fs 96320(1)$ & $11\degr55'55\farcs0800(3)$ & 21.9 & 270.0 \\
	bd179f1 & 2015.5699 & $15^{\rm h}37^{\rm m}09\fs 96319(1)$ & $11\degr55'55\farcs0787(5)$ & 14.4 & 245.7 \\
	bd192f0 & 2016.5875 &  $15^{\rm h}37^{\rm m}09\fs 96327(2)$ & $11\degr55'55\farcs0516(7)$ & 9.6 & 125.9 \\
	bd192f3 & 2017.1111 & $15^{\rm h}37^{\rm m}09\fs 96349(4)$ & $11\degr55'55\farcs0397(13)$  & 7.5 & 100.0 \\
	bd192f4 & 2017.1820 & $15^{\rm h}37^{\rm m}09\fs 96353(3)$ & $11\degr55'55\farcs0365(10)$  & 6.8 & 199.4 \\
	bd192f5 & 2017.2421 & $15^{\rm h}37^{\rm m}09\fs 96348(3)$ & $11\degr55'55\farcs0372(10)$  & 5.5 & 276.7 \\
	bd192f8 & 2017.5755 & $15^{\rm h}37^{\rm m}09\fs 96341(2)$ & $11\degr55'55\farcs0296(9)$  & 5.3 & 293.3 \\
	bd229a & 2020.6611 & $15^{\rm h}37^{\rm m}09\fs 96370(3)$ & $11\degr55'54\farcs9495(9)$  & 5.3 & 239.6 \\
	bd229b & 2020.6693 & $15^{\rm h}37^{\rm m}09\fs 96373(2)$ & $11\degr55'54\farcs9489(7)$  & 7.5 & 230.1 \\
	bd229c & 2021.0564 & $15^{\rm h}37^{\rm m}09\fs 96388(1)$ & $11\degr55'54\farcs9391(4)$  & 19.2 & 160.1 \\
	bd229d & 2021.0646 & $15^{\rm h}37^{\rm m}09\fs 96390(1)$ & $11\degr55'54\farcs9388(5)$  & 25.1 & 174.5 \\
	bd229e & 2021.4935 & $15^{\rm h}37^{\rm m}09\fs 96381(2)$ & $11\degr55'54\farcs9279(8)$  & 7.3 & 132.3 \\
	bd229f & 2021.5019 & $15^{\rm h}37^{\rm m}09\fs 96383(2)$ & $11\degr55'54\farcs9282(7)$  & 8.1 & 123.7 \\
	\hline
	\end{tabular}
    
    \tablenotetext{a}{In this table, the positional uncertainties have included both random and systematic errors (see Section~\ref{sec:results}). This table is available at \url{https://github.com/dingswin/publication_related_materials}, where the random errors for the positions can also be found.}
    \tablenotetext{b}{$(S/N)_\mathrm{J1537}$ and $(S/N)_\mathrm{SC}$ stand for the image S/N of (gated) J1537 and that of the secondary phase calibrator FIRST~J153746.2$+$114215, respectively.}
    
    \label{tab:pulsar_positions}
\end{table*}

\section{Astrometric results}
\label{sec:results}
Upon obtaining the 13 pulsar positions, we proceeded to estimate their systematic errors. 
This is because small residual calibration errors remain, even though direction-dependent calibration terms (of systematic errors) have been mitigated by the use of a close in-beam calibrator. 
We used the empirically derived expression from \citet{Deller19} to approach the systematic errors. For each epoch, the estimated systematic error was subsequently added in quadrature to the random error of the position. 
The positional uncertainties, including random and systematic errors, can be found in Table~\ref{tab:pulsar_positions} alongside the pulsar positions. To make it easier for other researchers to reproduce the error budget, the image S/N for both J1537 and the secondary phase calibrator are also presented in Table~\ref{tab:pulsar_positions}.
For the pulsar positions, the nominal systematic errors are around 0.14\,mas and 0.33\,mas in the right ascension (RA) and declination direction, respectively; in comparison, the median random errors are roughly twice the nominal systematic errors due to the faintness of J1537.

Based on the 13 pulsar positions and their
associated positional uncertainties (including the systematic errors described above),
we derived the astrometric results in three different methods: direct least-square fitting, bootstrap and Bayesian inference. 
Direct fitting was performed using {\tt pmpar} (\url{https://github.com/walterfb/pmpar}). A bootstrap was implemented as described in Section~3.1 of \citet{Ding20}. 
Compared to direct fitting and bootstrap, Bayesian analysis offers a simpler means to incorporate prior astrometric information (obtained elsewhere), and to infer extra orbital parameters (e.g., the longitude of ascending node and inclination angle) when positional precision allows \citep[e.g.][]{Deller13}.
We carried out Bayesian inference with {\tt sterne} (aStromeTry bayEsian infeReNcE, \url{https://github.com/dingswin/sterne}).
For the Bayesian inference, we assumed timing proper motion and parallax (reported in \citealp{Fonseca14}) follow Gaussian distributions, and used them as prior distributions for proper motion and parallax; the negligible (at the $5\,\mu$as level) reflex motion of J1537 (i.e., sky-position shifts due to the orbital motion) was not fitted.  
For both bootstrap and Bayesian analysis, we adopted the median value (of the marginalized sample) for an astrometric parameter as the estimate, and used the 16th and 84th percentiles to mark the $1\,\sigma$ uncertainty interval.

\begin{figure}
    \centering
	\includegraphics[width=9cm]{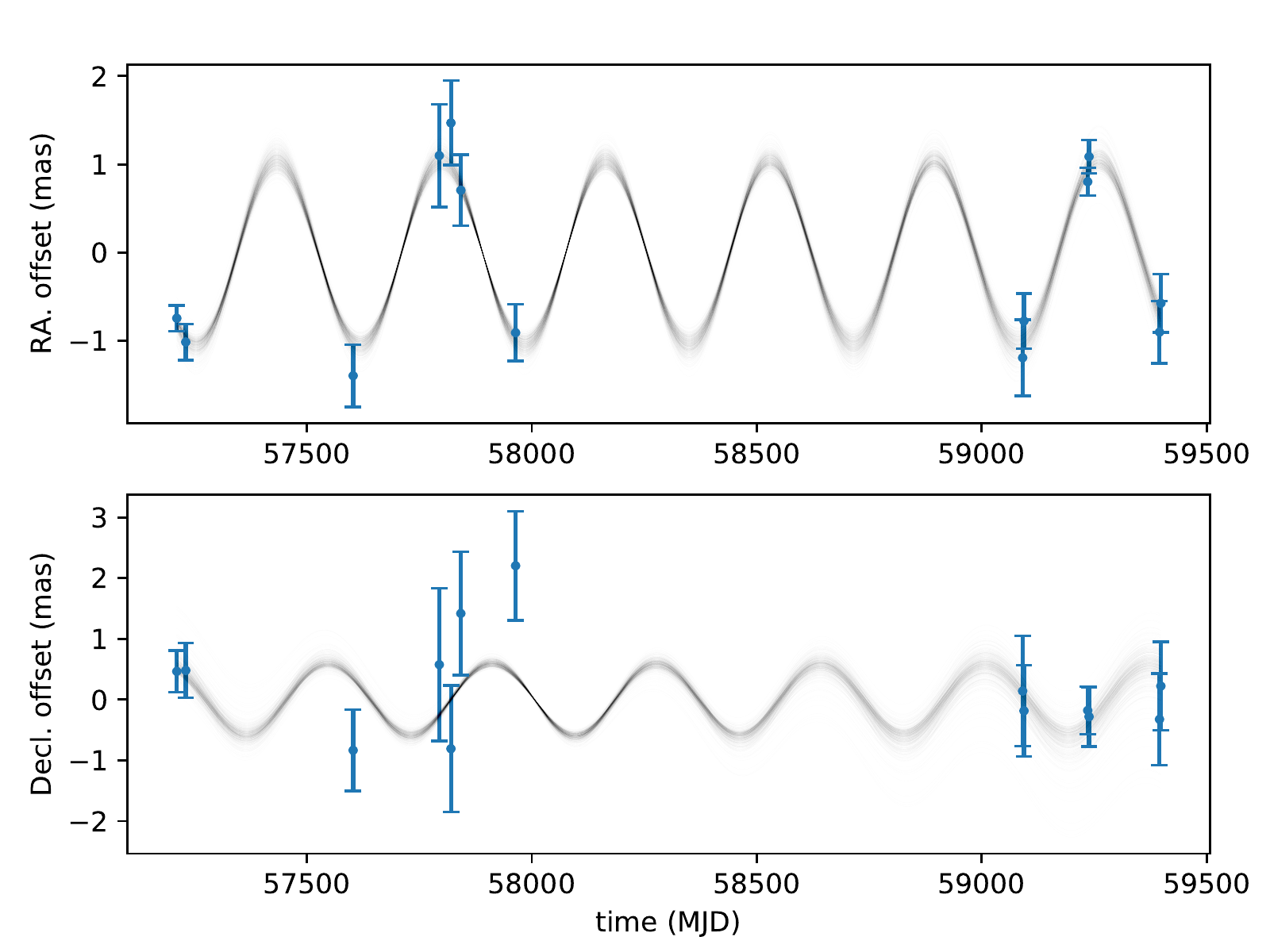}
    \caption{Parallax signature revealed by the \psr\ positions. Each quasi-sinusoidal curve represents the fitted model for a bootstrap run, after removing the best fit reference position and proper motion.}
    \label{fig:astrometric_model}
\end{figure}

\begin{figure*}
    \centering
	\includegraphics[width=12cm]{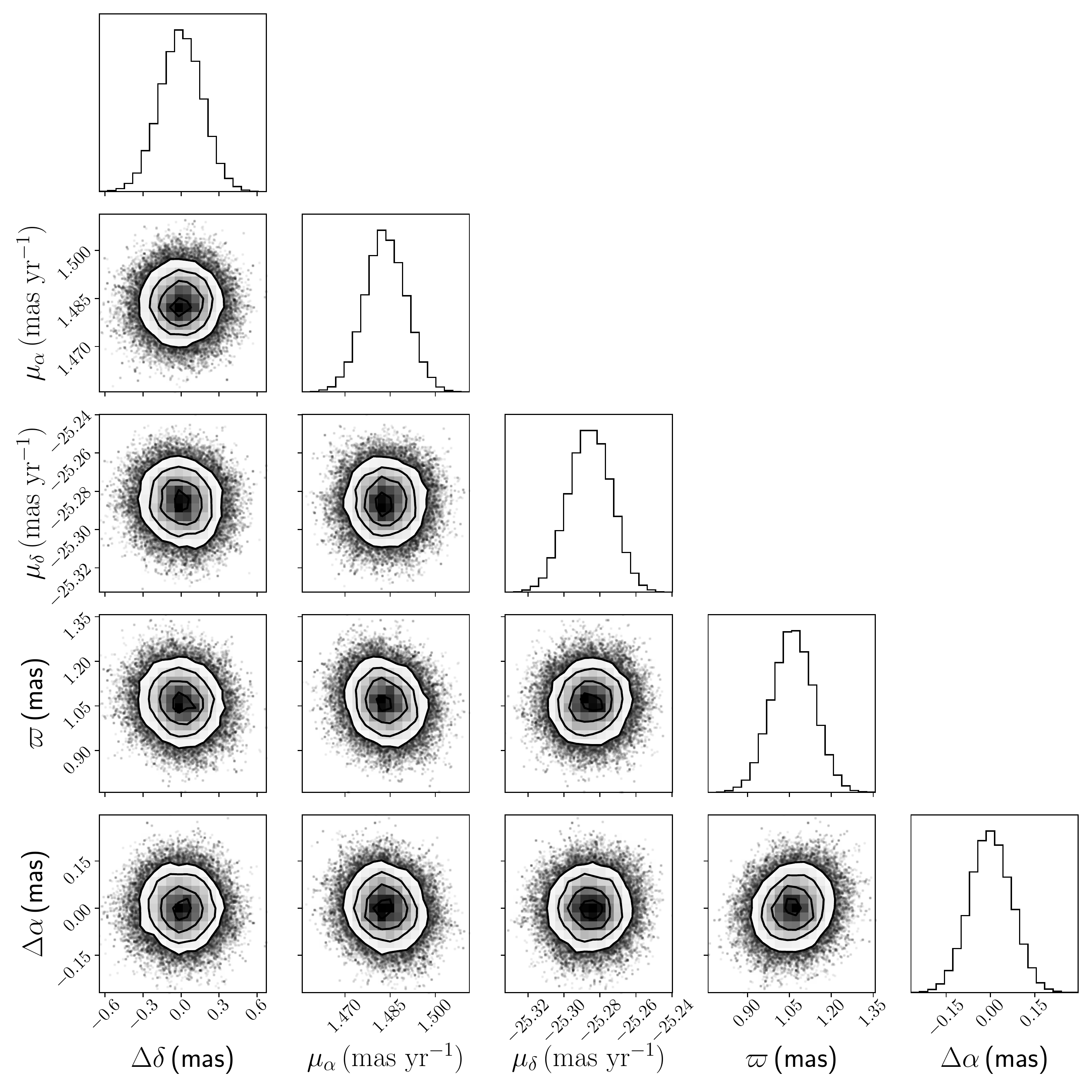}
    \caption{Error ``ellipses'' and the marginalized histograms for the posterior samples of the 5 astrometric parameters generated in the Bayesian analysis (see Section~\ref{sec:results}). 
    The reference position offset is relative to the median reference position provided in Table~\ref{tab:mu_and_pi}.}
    \label{fig:bayesian_cornerplot}
\end{figure*}

The astrometric results acquired with the three methods, as well as the three parallax-based distances, are summarized in Table~\ref{tab:mu_and_pi}. For comparison, the distances based on dispersion measure (DM) and pulsar timing are reproduced in Table~\ref{tab:mu_and_pi}. 
Here, the timing distance is quoted from the timing kinematic distance reported in \citet{Fonseca14}, which is derived from the orbital decay of J1537 by assuming GR is correct. In addition, the parallax signature, revealed by the 13 pulsar positions, is shown in Figure~\ref{fig:astrometric_model}. Figure~\ref{fig:bayesian_cornerplot} presents the posterior samples simulated with MCMC in the Bayesian analysis, which suggests negligible correlation in most (7 out of 10) pairs of astrometric parameters. However, small correlation is found between the three astrometric parameters having RA component, i.e., the reference RA $\alpha_\mathrm{J2000}$, the RA proper motion component $\mu_\alpha$ and the parallax $\varpi$. The largest correlation coefficient $|\rho|$ is 0.16 between $\varpi$ and $\alpha_\mathrm{J2000}$, while $|\rho|=0.14$ between $\mu_\alpha$ and $\varpi$.

According to Table~\ref{tab:mu_and_pi}, the astrometric results obtained with the three methods agree with each other; the new model-independent distances are generally consistent with the DM-based distance and the timing kinematic distance. The consistency between the new model-independent distances and the timing kinematic distance will be further improved in Section~\ref{sec:Pbdot_test} after updating the timing kinematic distance. The small reduced chi-square \rcs\ of 0.81 (for the method of direct fitting) implies that systematic errors for the 13 pulsar positions may have been slightly over-estimated. When applying the timing proper motion and parallax as prior information in the Bayesian analysis, \rcs\ only rises a little to 0.84, which indicates the timing proper motion and parallax \citep{Fonseca14} are consistent with the VLBI data.
Given a chi-square of $\sim17$ for 21 degrees of freedom, we did not see sufficient evidence to revise our estimated systematic uncertainties (reducing the estimated systematic uncertainty would bring the \rcs\ closer to unity and increase the parallax significance).

In the following discussion, we adopt the astrometric results derived with Bayesian analysis, which incorporates the VLBI and timing measurements. For those who want to use VLBI-only results (such as pulsar timers of J1537), we recommend the bootstrap results in Table~\ref{tab:mu_and_pi}, as bootstrap can potentially correct improper error estimations to an appropriate level (see \citealp{Ding20c} as a good example), especially when the number of measurements is relatively large ($\gtrsim10$ for VLBI astrometry).

\begin{table*}
    \raggedright
    \caption{Reference position, proper motion and parallax measurements of \psr\ at the reference epoch MJD~57964}
     
    	\begin{tabular}{cccccccc} 
		\hline
	method & $\alpha_\mathrm{J2000}$ (RA.) $^\mathrm{a}$ & $\delta_\mathrm{J2000}$ (Decl.) $^\mathrm{a}$ & $\mu_\alpha \equiv \dot{\alpha}\cos{\delta}$ & $\mu_\delta$ & $\varpi$ & \rcs\ & $D$  \\
	   & & & (\maspy) & (\maspy) & (mas) & & (kpc)  \\
		\hline
	direct fitting	& $15^{\rm h}37^{\rm m}09\fs 963467(4)$ & $11\degr55'55\farcs0274(1)$ & $1.51\pm0.02$ & $-25.31\pm0.05$ & $1.06\pm0.07$ & 0.81 & $0.94^{\,+0.07}_{\,-0.06}$ \\
	bootstrap & $15^{\rm h}37^{\rm m}09\fs 963467(4)$ & $11\degr55'55\farcs0274(2)$ & $1.51\pm0.02$ & $-25.31^{\,+0.04}_{\,-0.05}$ & $1.07^{\,+0.09}_{\,-0.08}$ & --- & $0.93\pm0.07$  \\
	Bayesian inference $^\mathrm{b}$ & $15^{\rm h}37^{\rm m}09\fs 963469(5)$ & $11\degr55'55\farcs0274(2)$ & $1.483\pm0.007$ & $-25.29\pm0.01$ & $1.063\pm0.075$ & 0.84 & $0.94^{\,+0.07}_{\,-0.06}$  \\
	\\
	dispersion measure & --- & --- & --- & --- & --- & --- & $0.7\pm0.2$ $^\mathrm{c}$  \\
	pulsar timing & --- & --- & $1.482\pm0.007$  & $-25.29\pm0.01$ & $0.86\pm0.18$ & --- & $1.051\pm0.005$ $^\mathrm{d}$ \\
	\hline
	\end{tabular}
    
    \tablenotetext{a}{All reference positions in this table only indicate the relative positions with respect to the second phrase calibrator. Accordingly, the reference position errors do not take into account the position errors of the main and second phrase calibrators.}
    \tablenotetext{b}{In the Bayesian analysis, we adopted the timing proper motion and parallax \citep{Fonseca14} as priors (assuming Gaussian distribution).}
    \tablenotetext{c}{\citet{Taylor93}. For the two newer Galactic free-electron distribution models \citep{Cordes02,Yao17}, timing-derived distances of J1537 have been incorporated into their establishment, thus becoming correlated to the associated DM-based distances.}
    \tablenotetext{d}{The timing results are reported in \citet{Fonseca14}; here, the quoted distance is the timing kinematic distance derived with the assumption that GR is correct (instead of with the timing parallax). We note that this timing kinematic distance is inferred with $\dot{P}_\mathrm{b}^{\,\mathrm{Gal}}=-3.5$\,\fsps\ (see Table~\ref{tab:Pbdot_Gal} and Section~\ref{sec:Pbdot_test} for more details) based on the Galactic mass distribution model by \citet{Nice95}.}
    
    \label{tab:mu_and_pi}
    \end{table*}

\section{Testing GR with the orbital decay of \psr}
\label{sec:Pbdot_test}
The observed orbital decay $\dot{P}_\mathrm{b}^\mathrm{obs}$ (or the observed time derivative of the orbital period) of J1537 has been estimated to be $-136.6\pm0.3$\,\fsps\ \citep{Fonseca14} from a global fit of the timing model, which can be attributed to
\begin{equation}
\label{eq:Pbdot_budget}
{\dot{P}_\mathrm{b}^{\,\mathrm{obs}}} = {\dot{P}_\mathrm{b}^{\,\mathrm{Gal}}} + {\dot{P}_\mathrm{b}^{\,\mathrm{Shk}}} + {\dot{P}_\mathrm{b}^{\,\mathrm{GW}}},
\end{equation}
where $\dot{P}_\mathrm{b}^{\,\mathrm{Gal}}$ and $\dot{P}_\mathrm{b}^{\,\mathrm{Shk}}$ stand for the extrinsic orbital decay due to the apparent effect of radial acceleration caused, respectively, by Galactic gravitational potential \citep{Damour91, Nice95} and by transverse motion \citep{Shklovskii70}; $\dot P_\mathrm{b}^{\,\mathrm{GW}}$ represents the intrinsic orbital decay as a result of the GW emissions from the inspiraling DNS. The estimation of the two extrinsic orbital-decay terms rely on the distance to J1537, while $\dot{P}_\mathrm{b}^{\,\mathrm{Shk}}$ also depends on the proper motion. 
On the other hand, the GR-based $\dot P_\mathrm{b}^{\,\mathrm{GW}}$ can be calculated, provided the orbital period $P_\mathrm{b}$, the orbital eccentricity $e$ and the masses of the two DNS constituents \citep{Peters63,Weisberg16}, all of which have been precisely determined with pulsar timing \citep{Fonseca14}. Hence, one can test GR by comparing the observed intrinsic orbital decay ${\dot{P}_\mathrm{b}^{\,\mathrm{int}}} =
{\dot{P}_\mathrm{b}^{\,\mathrm{obs}}} - {\dot{P}_\mathrm{b}^{\,\mathrm{Gal}}} - {\dot{P}_\mathrm{b}^{\,\mathrm{Shk}}}$ with $\dot P_\mathrm{b}^{\,\mathrm{GW}}$. For J1537, this test is the one (among the tests with PK parameters, see Section~\ref{subsec:DNS_pulsars}) that showed the largest discrepancy with GR (see Figure~9 of \citealp{Fonseca14}), possibly due to the unreliable DM-based distance used for the test.

Using Equation~22 of \citet{Weisberg16}, we calculated $\dot P_\mathrm{b}^{\,\mathrm{GW}}=-192.45\pm0.06$\,\fsps.
Using the proper motion $\mu=\sqrt{\mu_\alpha^2+\mu_\delta^2}$ and the distance $D$ acquired with Bayesian inference (see Table~\ref{tab:mu_and_pi}), we updated $\dot{P}_\mathrm{b}^{\,\mathrm{Shk}}=\mu^2 D/c \cdot P_\mathrm{b}=53\pm4$\,\fsps. 
The uncertainties for $\dot P_\mathrm{b}^{\,\mathrm{GW}}$ and $\dot{P}_\mathrm{b}^{\,\mathrm{Shk}}$ were derived with error propagation, which were subsequently confirmed by Monte-Carlo simulations.
In the $\dot{P}_\mathrm{b}^{\,\mathrm{Shk}}$ error estimation, we did not take into account the small correlation between $\mu_\alpha$ and $\varpi$ (mentioned in Section~\ref{sec:results}). This is because the correlation between $\mu$ and $\varpi$ is still negligible, as the declination component dominates the proper motion (see Table~\ref{tab:mu_and_pi}).

Following \citet{Zhu18}, we estimated $\dot{P}_\mathrm{b}^{\,\mathrm{Gal}}$ with different Galactic mass distribution models compiled in {\tt GalPot} (\url{https://github.com/PaulMcMillan-Astro/GalPot}, \citealp{McMillan17}).  
The $\dot{P}_\mathrm{b}^{\,\mathrm{Gal}}$ results are summarized in Table~\ref{tab:Pbdot_Gal}. 
The errors on $\dot{P}_\mathrm{b}^{\,\mathrm{Gal}}$ can be attributed to two sources: the uncertainty in the measurements (such as distance and proper motion) and the inaccuracy of the Galactic mass distribution model.
The former $\dot{P}_\mathrm{b}^{\,\mathrm{Gal}}$ errors, at the $\leq0.1$\,\fsps\ level (see Table~\ref{tab:Pbdot_Gal}), were derived with Monte-Carlo simulations.
We approached the latter $\dot{P}_\mathrm{b}^{\,\mathrm{Gal}}$ errors with the standard deviation of the $\dot{P}_\mathrm{b}^{\,\mathrm{Gal}}$ estimates listed in Table~\ref{tab:Pbdot_Gal}. For this calculation of the standard deviation, we do not include the $\dot{P}_\mathrm{b}^{\,\mathrm{Gal}}$ based on the analytical model by \citet{Nice95}, because {\bf 1)} the analytical model is oversimplified (see the discussion in Appendix~A of \citealp{Zhu18}) and {\bf 2)} the resultant $\dot{P}_\mathrm{b}^{\,\mathrm{Gal}}$ is inconsistent with other models (see Table~\ref{tab:Pbdot_Gal}). Accordingly, we adopted the average $\dot{P}_\mathrm{b}^{\,\mathrm{Gal}}$ of the four remaining Galactic mass distribution models \citep{Dehnen98,Binney11,Piffl14,McMillan17} as the $\dot{P}_\mathrm{b}^{\,\mathrm{Gal}}$ estimate. 
In this way, we obtained $\dot{P}_\mathrm{b}^{\,\mathrm{Gal}}=-1.9\pm0.2$\,\fsps, where the error budget has included the standard deviation (0.14\,\fsps) of $\dot{P}_\mathrm{b}^{\,\mathrm{Gal}}$. 
As {\tt Galpot} was not available in 2014, \citet{Fonseca14} adopted the $\dot{P}_\mathrm{b}^{\,\mathrm{Gal}}$ based on the Galactic mass distribution model of \citet{Nice95} (see Table~\ref{tab:Pbdot_Gal}) for the calculation of the timing kinematic distance. Provided our new $\dot{P}_\mathrm{b}^{\,\mathrm{Gal}}$, the timing kinematic distance of $1.05$\,kpc (reported by \citealp{Fonseca14} and quoted in Table~\ref{tab:mu_and_pi}) would decrease by 3\% to $1.02$\,kpc, thus becoming consistent with the new model-independent distance (see Table~\ref{tab:mu_and_pi}).

\begin{table}
    \raggedright
    \caption{Galactic-potential-related orbital decay $\dot{P}_\mathrm{b}^{\,\mathrm{Gal}}$ as well as its two components (i.e., $\dot{P}_\mathrm{b,h}^{\,\mathrm{Gal}}$ and $\dot{P}_\mathrm{b,z}^{\,\mathrm{Gal}}$, respectively corresponding to the component horizontal and vertical to the Galactic plane) estimated with different models of Galactic mass distribution. The $\dot{P}_\mathrm{b}^{\,\mathrm{Gal}}$ uncertainties for the models are derived with Monte-Carlo simulations. For comparison, the $\dot{P}_\mathrm{b}^{\,\mathrm{Gal}}$ expected by GR (i.e., ${\dot{P}_\mathrm{b}^{\,\mathrm{obs}}} - {\dot{P}_\mathrm{b}^{\,\mathrm{Shk}}} - {\dot{P}_\mathrm{b}^{\,\mathrm{GW}}}$) is provided.}
    
       	\begin{tabular}{cccc} 
		\hline
	Galactic mass & $\dot{P}_\mathrm{b,h}^{\,\mathrm{Gal}}$ & $\dot{P}_\mathrm{b,z}^{\,\mathrm{Gal}}$ & $\dot{P}_\mathrm{b}^{\,\mathrm{Gal}}$  \\
	distribution model & (\fsps) & (\fsps) & (\fsps) \\
		\hline
	\citet{Nice95} & $1.1(1)$ & $-4.6(1)$ & $-3.51(6)^\mathrm{a}$ \\
	\citet{Dehnen98} & $0.120(1)$ & $-2.04(6)$ & $-1.92(6)^\mathrm{b}$  \\
	\citet{Binney11} & $0.131(3)$ & $-1.89(9)$ & $-1.76(9)$ \\
	\citet{Piffl14} & $0.132(1)$ & $-2.09(8)$ & $-1.96(8)$ \\
	\citet{McMillan17} & $0.141(2)$ & $-2.2(1)$ & $-2.1(1)$\\
	\hline
	${\dot{P}_\mathrm{b}^{\,\mathrm{obs}}} - {\dot{P}_\mathrm{b}^{\,\mathrm{Shk}}} - {\dot{P}_\mathrm{b}^{\,\mathrm{GW}}}$ & --- & --- & $2.5(3.8)$ \\
	
	\hline
	\end{tabular}
	\tablenotetext{a}{For the calculation, we adopted $R_0=8.12\pm0.03$\,kpc (the distance from the Sun to the Galactic center) provided by \citet{Gravity-Collaboration18} and $\Theta_0=234.6\pm1.1\,{\rm km~s^{-1}}$ (the circular speed of the local standard of rest). We derived the $\Theta_0$ with the proper motion of Sgr~A* \citep{Reid20}, the aforementioned $R_0$ \citep{Gravity-Collaboration18} and the velocity of the Sun with respect to the local standard of rest \citep{Schonrich10}.}
    \tablenotetext{b}{There are 4 models discussed in \citet{Dehnen98}. Here, we used the ``model 3'', which falls into the middle of the models 1 to 4, and is generally consistent with the other 3 models.}
    \label{tab:Pbdot_Gal}
\end{table}

\begin{figure}
    \centering
    \hspace*{-0.5cm}
	\includegraphics[width=11cm]{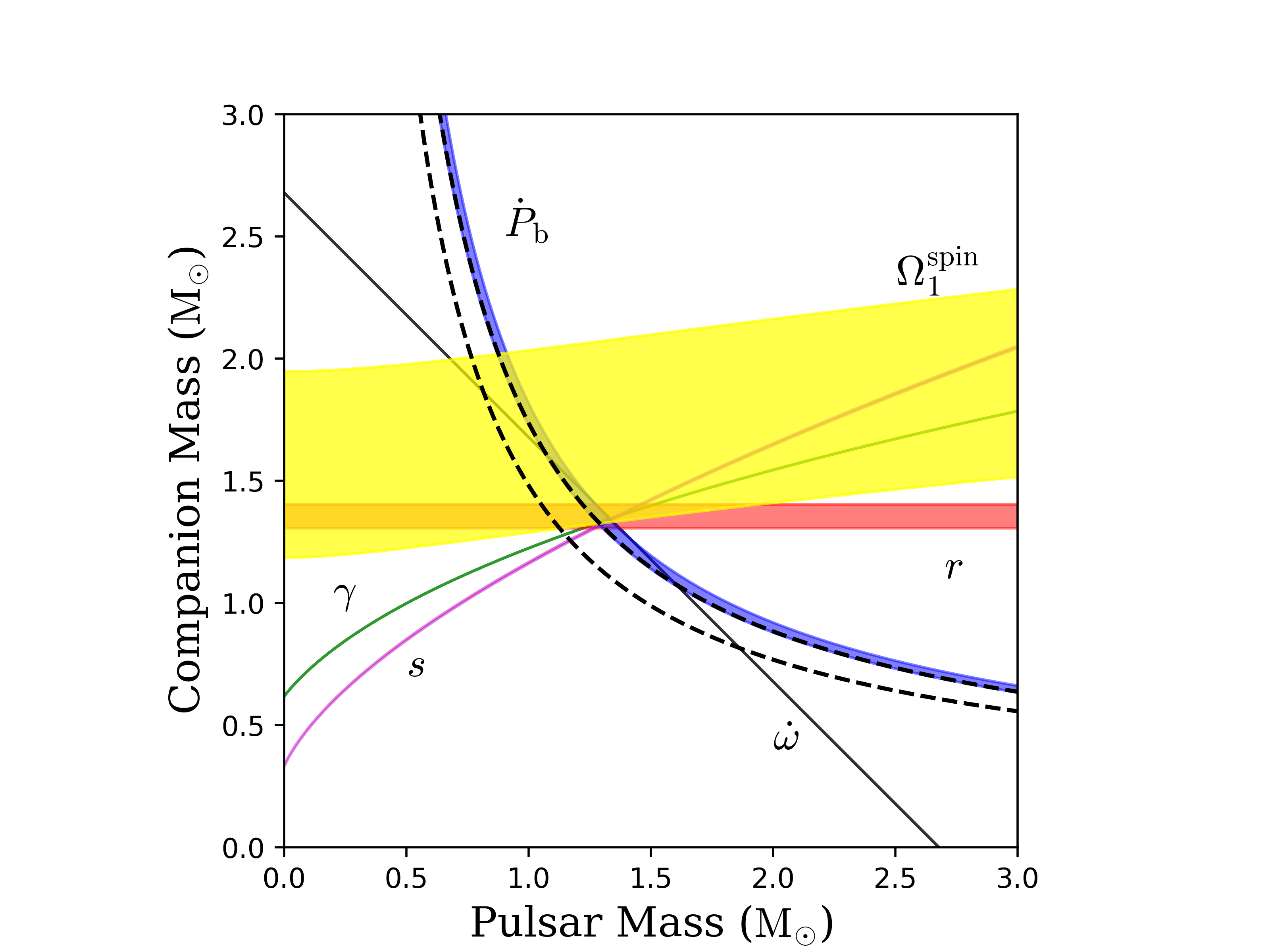}
    \caption{Mass-mass diagram of J1537. Its only difference from Figure~9 of \citet{Fonseca14} is the updated mass-mass constraint offered by the new observed intrinsic orbital decay ${\dot{P}_\mathrm{b}^{\,\mathrm{int}}}$, shown with the blue strip. For comparison, the mass-mass constraint given by the previous ${\dot{P}_\mathrm{b}^{\,\mathrm{int}}}=-0.17(1)\,\mathrm{ps~s^{-1}}$ inferred from the DM-based distance (see Table~\ref{tab:mu_and_pi}) is provided with two dashed curves.
    For the other PK parameters, the green, pink, red, black and yellow strips stand for the mass-mass constraints posed by the time-averaged gravitational redshift $\gamma=2.0708(5)$\,ms, the Shapiro delay ``shape'' $s=0.977(2)$, the Shapiro delay ``range'' $r=6.6(2)\,\mu$s, the periastron advance rate $\dot{\omega}=1.755795(2)\,\mathrm{deg~{yr}^{-1}}$ and the pulsar precession rate $\Omega_1^\mathrm{spin}=0.59^{+0.12}_{-0.08}\,\mathrm{deg~{yr}^{-1}}$, respectively \citep{Fonseca14}.}
    \label{fig:mass_mass_diagram}
\end{figure}

Collectively, we reached ${\dot{P}_\mathrm{b}^{\,\mathrm{int}}}=-188.0\pm3.8$\,\fsps, corresponding to 
\begin{equation}
\label{eq:Pbdot_test_result}
\frac{\dot{P}_\mathrm{b}^{\,\mathrm{int}}}{\dot{P}_\mathrm{b}^{\,\mathrm{GW}}}=0.977\pm0.020 \,,
\end{equation}
which is the third most precise orbital-decay test of GR in the strong-field regime according to Table~3 of \citet{Weisberg16}. 
At the 2\% precision level, the new observed intrinsic orbital decay is within $1.2\,\sigma$ of the GR prediction (see Figure~\ref{fig:mass_mass_diagram}), which relieves the mild tension of the previous $\dot{P}_\mathrm{b}$ test (${\dot{P}_\mathrm{b}^{\,\mathrm{int}}}/{\dot{P}_\mathrm{b}^{\,\mathrm{GW}}}=0.91\pm0.06$ at 1.7\,$\sigma$ agreement, \citealp{Stairs02}). 

For visualisation, the mass-mass diagram of J1537 (updated from Figure~9 of \citealp{Fonseca14}) is presented in Figure~\ref{fig:mass_mass_diagram}, which involves 6 PK parameters. 
Apart from the 5 PK parameters already mentioned in Section~\ref{sec:intro}, $\Omega_1^\mathrm{spin}$ stands for the precession rate of the pulsar. 
Each PK parameter is a function of the two DNS constituent masses. Therefore, each observed PK parameter (and its uncertainty) offers a constraint on the two masses. If GR is correct, all mass-mass constraints should converge at the ``true'' masses of the pulsar and its companion. In Figure~\ref{fig:mass_mass_diagram}, this convergence is visible with the new ${\dot{P}_\mathrm{b}^{\,\mathrm{int}}}$.

Looking into the future, the bottleneck of the orbital-decay test with J1537 will continue to be its parallax uncertainty, which would decrease with $t^{-1/2}$ (e.g. \citealp{Ding21}; here, $t$ stands for the on-source time instead of the time span) in the long term despite fluctuations of J1537 brightness. This process of precision enhancement would be accelerated with high-sensitivity VLBI observations, as the VLBA observations of J1537 are generally sensitivity-limited, especially when J1537 is down-scintillated.

\section*{Acknowledgements}
We are grateful to the anonymous referee for the swift and valuable comments, and appreciate the helpful discussions with Leonid Petrov, David Kaplan and Joseph Lazio. B.S. and A.L. thank Mitch Mickaliger for help with timing data acquisition and preparation.
H.D. is supported by the ACAMAR (Australia-ChinA ConsortiuM for Astrophysical Research) scholarship, which is partly funded by the China Scholarship Council (CSC).
A.T.D is the recipient of an ARC Future Fellowship (FT150100415).
Pulsar research at UBC is supported by an NSERC Discovery Grant and by the Canadian Institute of Advanced Research.
Pulsar research at JBCA is supported by a Consolidated Grant from STFC. 
Parts of this research were conducted by the Australian Research Council Centre of Excellence for Gravitational Wave Discovery (OzGrav), through project number CE170100004.
This work is based on observations with the Very Long Baseline Array (VLBA), which is operated by the National Radio Astronomy Observatory (NRAO). The NRAO is a facility of the National Science Foundation operated under cooperative agreement by Associated Universities, Inc.
Data reduction and analysis was performed on OzSTAR, the Swinburne-based supercomputer.
This work made use of the Swinburne University of Technology software correlator, developed as part of the Australian Major National Research Facilities Programme and operated under license. 


\bibliography{haoding}{}
\bibliographystyle{aasjournal}



\end{document}